\documentclass[josa,twocolumn,showpacs,floatfix,superscriptaddress,10pt]{revtex4-1}

\usepackage{amssymb}
\usepackage[usenames,dvipsnames]{color}
\usepackage{graphicx}
\usepackage{subfigure}
\usepackage{epsfig}
\usepackage{amsmath}
\usepackage{subfigure}
\usepackage[normalem]{ulem}

\begin{document}

\title{An experimental proposal for a Gaussian amendable quantum channel}

\author{D. Buono}
\affiliation{Dipartimento di Ingegneria Industriale, Universit\`{a} degli Studi di
Salerno, via Giovanni Paolo II, I-84084 Fisciano (SA), Italy}
\author{G. Nocerino}
\affiliation{Trenitalia spa, DPR Campania, Ufficio di Ingegneria della Manutenzione, IMC
Campi Flegrei, Via Diocleziano 255, 80124 Napoli, Italy}
\author{A. Porzio}
\affiliation{CNR - SPIN, Napoli, Complesso Universitario Monte Sant'Angelo, I-80126
Napoli, Italy}
\author{A. Mari}
\affiliation{NEST, Scuola Normale Superiore and Istituto Nanoscienze-CNR, \\
piazza dei Cavalieri 7, I-56126 Pisa, Italy}
\author{A. De Pasquale}
\affiliation{NEST, Scuola Normale Superiore and Istituto Nanoscienze-CNR, \\
piazza dei Cavalieri 7, I-56126 Pisa, Italy}
\author{V. Giovannetti}
\affiliation{NEST, Scuola Normale Superiore and Istituto Nanoscienze-CNR, \\
piazza dei Cavalieri 7, I-56126 Pisa, Italy}

\begin{abstract}
We propose a quantum optics experiment where a single two-mode
Gaussian entangled state is used for realizing the paradigm of an amendable
Gaussian channel recently presented in Phys. Rev. A, \textbf{87}, 062307 (2013).
Depending on the choice of the experimental parameters the entanglement
of the probe state is preserved or not and the relative map belongs or
not to the class of entanglement breaking channels.
The scheme has been optimized to be as simple as possible: it requires only a single active non-linear operation followed by four passive beam-splitters. The effects of
losses, detection inefficiencies and statistical errors are also taken into
account, proving  the feasibility of the experiment with current realistic resources.
\end{abstract}

\maketitle

\section{Introduction}

Decoherence embodies the
detrimental effects of noise on any quantum system whose coherence, in its
widest sense, is smeared causing the loss of information of the
initial state \cite{zurek}.
This represents a focal point in quantum information theory  \cite{nc} as it limits both the attainable fidelity and the variety of accessible protocols \cite{zurek,deco,PRA2012}. 
In particular, entanglement~\cite{REVENT01,REVENT02} represents a fundamental resource in quantum computation \cite{nc} and thus it 
should be protected against dechorence.
In this regard, the most  \textquotedblleft undesirable\textquotedblright family of quantum processes is given by the so-called entanglement breaking (EB) maps \cite{EBT,holevoEBG01,holevoEBG02,holevoEBG03}, under whose action any entanglement initially installed between the system and an external ancilla is completely lost.
These are maps acting on one component of an entangled pair leaving unperturbed the other.

\textit{Amendable channels} strongly related to EB maps \cite{mucritico}.
They realize an EB map when applied twice consecutively on the same system, but they admit a {\it filtering} operation that, 
applied in between the 
first and the second action of the map, prevents the global transformation from
being entanglement breaking.
Identifying the set of amendable channels and their associated 
filtering operations is an important quantum error correction task which have profound implications in many research areas.
In particular this 
could be useful in developing 
efficient long-range communication schemes based on quantum repeaters architectures~\cite{QREP01,QREP02,QREP03} where the signaling
process takes place  through intermediaries (the  quantum repeaters) who collect, process, and  redistribute the messages sent by the communicating parties (in this picture
the action of an amendable channel simulates the transferring from two communicating  parties and one repeater, while the filtering operation corresponds to the data processing performed by the latter). 

The study of amendable channels is particularly 
relevant in the context of the so called Bosonic Gaussian Channels (BGCs)~\cite{gaussian01,gaussian02,gaussian03,gaussian04}.
These are completely positive trace preserving maps \cite{deco,review}, 
which provide prototypical examples of decoherence processes  that occurs in  continuous variable (CV) systems~\cite{CVSYS}, e.g. in the transmission of optical signals 
through lossy dispersive optical fibers and/or in free-space~ \cite{REVCAVES}.
Examples of BGCs which are 
amendable were first discussed in Ref~\cite{amendableChTEO}. Moving from those observations, in this paper we propose and discuss in details a feasible quantum
optics experiment for the realization and the experimental test of an amendable map using Gaussian
channels. In particular, having at disposal a two-mode
squeezed vacuum state \cite{PRL2009} generated by a type-II sub--threshold OPO \cite{APB2008}, 
we show that by suitable passive linear optical manipulations 
it is possible to realize an EB Gaussian channel. Then
we prove that it is possible to amend the EB channel in a simple way thus preserving the initial entanglement of the probe state.
The proposed experimental set--up is an effective realization of the
conceptual scheme discussed in Sec. III A of Ref. \cite{amendableChTEO}.

The paper is structured as follows: in Sec. \ref{Sect:Review} we present a
brief review of the theory of Gaussian amendable channels introducing some useful
notation (see \ref{subSect:notation}) and the conceptual theoretical scheme
(see \ref{subSect:teoscheme}). In Sec. \ref{Sect:sketch} we give a glance
over the experimental proposal. In particular we prove that (see \ref{subSect:expI})
a proper manipulation of the output of a single type--II
sub--threshold OPO is sufficient for generating both an entangled probe state and
a local squeezed ancilla. Then, we show that an effective EB channel can be obtained
by using only passive optical elements (see \ref{subSect:expII}). In Sec. \ref{Sect:Results},  we
estimate the correlations of the output state looking for suitable experimental conditions that would make EB
the resulting map. Then, we find the parameters setting that makes the map
effectively amendable.  Finally, (see \ref{subSect:losses}) we analyze the feasibility
of the experiment in presence of losses, measurement uncertainty and detection inefficiencies.

\section{Review of the theory of Gaussian amendable channels \label{Sect:Review}}

In this section we review some basic theoretical notions and discuss a
simple example of Gaussian amendable channel. A more detailed analysis can
be found in Ref.\ \cite{amendableChTEO}.

\subsection{Notation\label{subSect:notation}}

Consider $n$ optical radiation modes described by their
position and momentum quadrature operators $q_{1}$, $q_2$,
$\cdots$, $q_n$,  and $p_{1}$, $p_2$, $\cdots$, $p_n$ which we group in a vector of
 $2n$ components: $R=(q_{1},p_{1},\dots q_{n},p_{n})$. Such operators can be chosen to be dimensionless so that they
obey the canonical commutation rules $[q_{i},p_{j}]=i\delta _{i,j}$, $[q_{i},q_{j}]=[p_{i},p_{j}]=0$. 
To any state $\rho$ of the system
we can associate its first and second statistical moments
defined respectively by the real vector $\langle R\rangle$
and by the $2n\times 2n$ covariance matrix (CM) $V$ with entries 
\begin{equation}
V_{ij}=\frac{\langle R_{i}R_{j}+R_{j}R_{i}\rangle }{2}-\langle R_{i}\rangle
\langle R_{j}\rangle ,
\end{equation}%
where the symbol 
$\langle \cdots \rangle$ indicates expectation values
with respect to $\rho$. 
Gaussian density matrices are fully characterized once
$\langle R\rangle$ and $V$ are assigned~\cite{HOLEVOBOOK}.
 They correspond to states of the $n$--mode system  whose
 associated characteristic function is Gaussian.
Examples of Gaussian states which will play an important role in
the next section are the following pure states: single mode vacuum state, single mode squeezed vacuum state and two-mode squeezed vacuum state (TMSV). According to this notation the vacuum
state is characterized by $\langle R\rangle =(0,0)$ and 
\begin{equation}
V_{0}=\frac{1}{2}\left( 
\begin{array}{cc}
1 & 0 \\ 
0 & 1%
\end{array}%
\right) ,  \label{vacuum}
\end{equation}
the squeezed state by $\langle R\rangle =(0,0)$ and 
\begin{equation}
V_{1}(r)=\frac{1}{2}\left( 
\begin{array}{cc}
e^{r} & 0 \\ 
0 & e^{-r}%
\end{array}%
\right) ,  \label{squeezed}
\end{equation}%
while the TMSV state has $\langle R\rangle =(0,0,0,0)$ and 
\begin{equation}
V_{2}(r)=\frac{1}{2}\left( 
\begin{array}{cccc}
\cosh (r) & 0 & \sinh (r) & 0 \\ 
0 & \cosh (r) & 0 & -\sinh (r) \\ 
\sinh (r) & 0 & \cosh (r) & 0 \\ 
0 & -\sinh (r) & 0 & \cosh (r)%
\end{array}%
\right) .  \label{TMSV}
\end{equation}%
Gaussian channels are quantum operations which map Gaussian states into
Gaussian states~\cite{gaussian01,gaussian02,gaussian03,gaussian04}. Therefore, they are completely defined by their action on
the displacement vector $\langle R\rangle $ and the matrix $V$. Moreover, since the
level of entanglement of a state depends only on the correlations and it is
insensitive to displacement operations, 
the action on $\langle R\rangle$ can be completely neglected for the  purpose of the present paper. 
In particular in the following we will make extensive use of the transformation associated to a
beam splitter of transmissivity $\eta$. Given two input modes
with CM $V$ it will produce 
at the output a two-mode state with CM $V^{\prime }(\eta )=B(\eta )VB(\eta )$, where 
\begin{equation}
B(\eta )=\left( 
\begin{array}{cccc}
\sqrt{\eta } & 0 & \sqrt{1-\eta } & 0 \\ 
0 & \sqrt{\eta } & 0 & \sqrt{1-\eta } \\ 
\sqrt{1-\eta } & 0 & -\sqrt{\eta } & 0 \\ 
0 & \sqrt{1-\eta } & 0 & -\sqrt{\eta }%
\end{array}%
\right) .  \label{bs}
\end{equation}%
If we mix a single mode state with the vacuum on a beam splitter and we
trace out one of the output modes we are left with a non-unitary attenuation (or lossy) 
channel $\Phi _{At}(\eta )$~\cite{CGH}  acting on the CMs as 
\begin{equation}
V\rightarrow V^{\prime }=\eta V+(1-\eta )V_{0},  \label{attenuation}
\end{equation}%
where $V_{0}$ is the CM of the vacuum given in Eq.\ (\ref{vacuum}). Another
important single mode operation we will use in the following is the single mode squeezing
acting as $V\rightarrow V^{\prime }(r)=S(r)VS(r)$ with
\begin{equation}
S(r)=\left( 
\begin{array}{cc}
e^{r} & 0 \\ 
0 & e^{-r}%
\end{array}%
\right) .  \label{squeezing}
\end{equation}%
The previous states, operations and combinations thereof are the main
ingredients of the scheme which will be presented in the following. Finally we stress 
that in a real experiment the CM of, at most, a two--mode state can be fully reconstructed by a
single homodyne detection scheme \cite{JOB2005}.

\subsection{Theoretical scheme\label{subSect:teoscheme}}

Our goal is identifying an experimentally feasible scheme for realizing the paradigms of Gaussian amendable channels.
As recalled in the introduction a channel $%
\Phi $ is amendable if it is entanglement breaking of order $2$, i.e. 
\begin{equation}
\Phi \circ \Phi \in \mathrm{EB},  \label{amend1}
\end{equation}
and there exists a unitary filter such that 
\begin{equation}
\Phi \circ \mathcal{U}\circ \Phi \notin \mathrm{EB}.  \label{amend2}
\end{equation}
This problem is equivalent to the
following one: find a channel $\Phi ^{\prime }$ and a unitary $\mathcal{U}%
^{\prime }$ such that 
\begin{equation}  \label{amend3}
\Phi ^{\prime }\circ \mathcal{U}^{\prime }\circ \Phi ^{\prime }\in \mathrm{EB%
},
\end{equation}
while 
\begin{equation}  \label{amend4}
\Phi ^{\prime }\circ \Phi ^{\prime }\notin \mathrm{EB}.
\end{equation}
Indeed if Eq.s (\ref{amend3}) and (\ref{amend4}) hold, it is straightforward to
check that $\Phi =\mathcal{U}\circ \Phi ^{\prime }$ and $\mathcal{U}=%
\mathcal{U}^{\prime \dag }$ satisfy Eq.s (\ref{amend1}) and (\ref{amend2}) in view of the invariance of entanglement under local unitaries.
It turns out that, for Gaussian channels, the second problem is simpler to
address and therefore in this paper we focus on the latter pair of conditions, Eq.s (\ref{amend3},\ref{amend4}).

In Ref. \cite{amendableChTEO}, it was shown that an
attenuation channel (Eq.\ (\ref{attenuation})) and a local squeezing
operation (Eq.\ (\ref{squeezing})) are valid examples of $\Phi^{\prime}$ and 
$\mathcal{U}^{\prime}$ respectively. Indeed one has that, for some values of 
the channel transmissivity $\eta$ and squeezing parameter $r$
\begin{equation}
\Phi_1 := \Phi_{At}(\eta) \circ \mathcal{S}(r) \circ \Phi_{At}(\eta) \in {%
\mathrm{EB}} ,  \label{phi1}
\end{equation}
while 
\begin{equation}
\Phi_2 := \Phi_{At}(\eta) \circ \Phi_{At}(\eta) \notin {\mathrm{EB}} .
\label{phi2}
\end{equation}

A natural way to verify that $\Phi _{1}$ is $\mathrm{EB}$ while $\Phi _{2}$
is not would be to apply those maps to one part of a maximally entangled state
and check whether the initial entanglement is preserved or not.
In continuous variables systems, however maximally entangled states are
not physically realizable but, as proven in Ref. \cite{amendableChTEO},
the test can be performed by using a pure two-mode squeezed state with finite
entanglement and mean energy. However, we note here that if the incoming state
is mixed, e.g. due to the presence of losses in the state preparation stage,
this equivalence property is not valid any more and the output state may result
in being separable even if the map is not EB. It goes without saying that even in
this case it may happen that a unitary filter, acting at a proper stage,
will restore the lost entanglement.

\begin{figure}[t]
\includegraphics[width=0.9 \columnwidth]{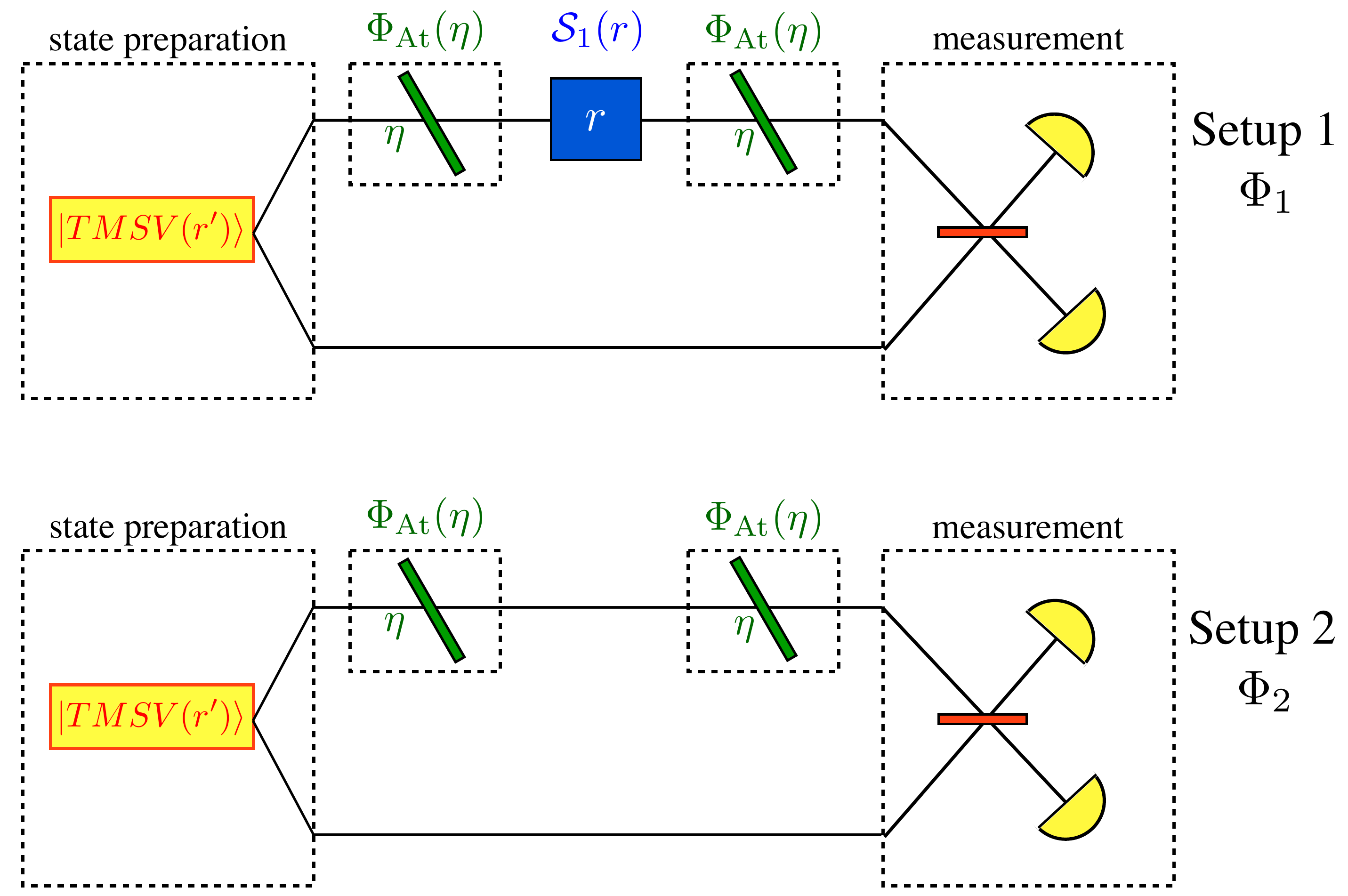}
\caption{(Color online) Theoretical scheme of a specific example of channels
satisfying Eq.s (\protect\ref{amend3}) and (\protect\ref{amend4}) proving
the existence of Gaussian amendable maps. In both cases a TMSV state is
prepared. In Setup 1 the channel $\Phi _{1}$ is applied to one part of the
state, while in setup 2 the channel $\Phi _{2}$ is used. By measuring the
entanglement of output state one can check for which set of parameters $\Phi
_{1}$ is EB while $\Phi _{2}$ is not EB.}
\label{fig:simplifiedsetups}
\end{figure}

The theoretical model of this test is graphically depicted in Fig.\ \ref%
{fig:simplifiedsetups}, where we sketch the scheme of $\Phi_1$ and $\Phi_2$. In both cases the map is applied to one side
of a TMSV state with squeezing parameter $r^{\prime}$ (see Eq.\ (\ref{TMSV})).
After the application of $\Phi_1$ it has been proved (see Fig. 4a of Ref. \cite{amendableChTEO}) that there exists a range of $\eta$ for which the output state
is separable. On the other hand, when $\Phi_2$ is applied the output state is
always entangled, for all values of $r^{\prime}$ and $\eta$ (see Fig. 4b of Ref. \cite{amendableChTEO}).
The equivalence between Eq.s (\ref{amend3},\ref{amend4}) and
Eq.s (\ref{amend1},\ref{amend2}) implies that the Gaussian map 
$\Phi=\mathcal{S}(r) \circ \Phi_{At}(\eta)$ is amendable via a squeezing
unitary filter $\mathcal{S}^\dag(r)$. This theoretical scheme will be our
starting point for designing a more realistic experimental proposal.

\section{Experimental proposal\label{Sect:sketch}}

In this section we propose an experimental set--up in order to realize an
effective amendable Gaussian map, endowed with the appealing property of
being quite simple to be realized in the laboratory. Furthermore, we will
also take into account the effects of losses, detection inefficiency and,
eventually, measurement indeterminacy.

\subsection{Preparation stage\label{subSect:expI}}

\begin{figure}[!ht]
\begin{center}
\includegraphics[width=0.6\columnwidth]{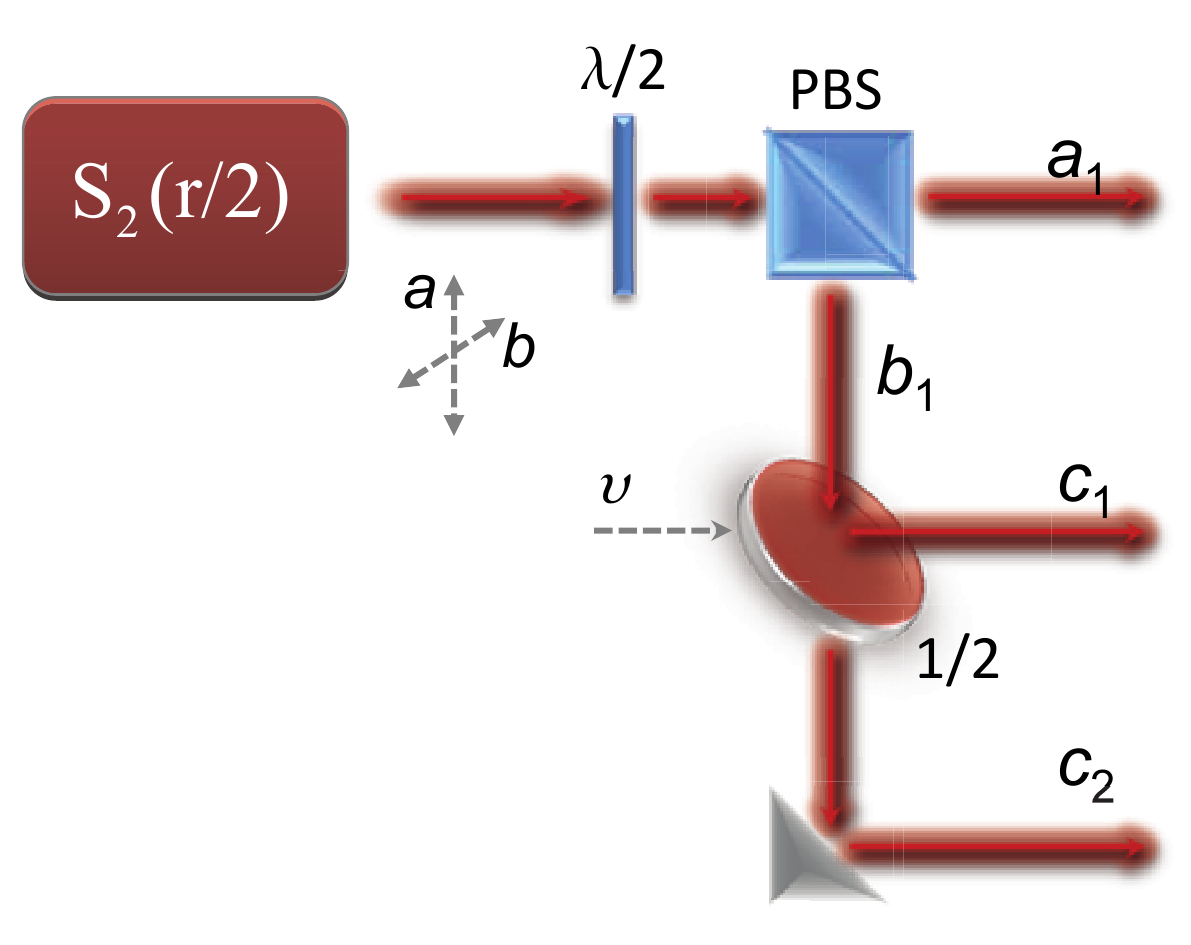}
\end{center}
\caption{(Color online) Experimental scheme for the generation of the
resource states. A type II OPO gives a twin beam described by the covariance
matrix (\protect\ref{TMSV}). The entangled modes $a$ and $b$ are orthogonally
polarized. By applying a $\protect\lambda /2$ wave plate and a polarizing
beam splitter (PBS) to $a$ and $b$, we can obtain two \textit{independent}
single mode squeezed vacuum states $a_{1}$ and $b_{1}$ with orthogonal
squeezing phases (Eq.~(\protect\ref{resource1})). Then, through a
balanced beam splitter (the red one), polarization insensitive, we mix mode $%
b_{1}$ with the vacuum state $v$ obtaining the pair $\left(
c_{1},c_{2}\right)$, a pure TMSV (Eq.~(\ref{eq:Vc1c2})). This pair represents the entangled system
on which we will test the entanglement--breaking properties of the maps $\Phi
_{1}$ and $\Phi _{2}$ defined in Eq.s (\protect\ref{phi1},\protect\ref{phi2}).
The pure single mode beam $a_{1}$ will be used for implementing the
squeezing transformation $\mathcal{S}(r)$.}
\label{FigurePartScheme}
\end{figure}

Our starting point is the theoretical model given in Fig.\ \ref%
{fig:simplifiedsetups}. Notice that, even though appearing quite simple, the
first set--up associated with the channel $\Phi _{1}$ in principle
requires, in addition to the state preparation part, a non-trivial
active operation on the system: the local squeezing between the two beam
splitters. In quantum optics active operations can be realized by non-linear interactions. In most squeezing schemes, the initial state (e.g. vacuum,
coherent, squeezed and/or thermal) interacts with a strong classical field
in an optical nonlinear medium. This can be achieved for example 
by a sub--threshold
optical parametric oscillator (OPO) \cite{squeezedOPO}. Compared with passive transformations,
active operations are relatively difficult to be engineered and
experimentally costly. 
In order to get around this obstacle, our idea is to use a single initial
active operation both for the generation of the entangled (probe) state and for
the (indirect) realization of the local squeezing. These resources will be
obtained in the preparation stage described in Fig.\ \ref{FigurePartScheme}.

At the output of a type II OPO one has at disposal two cross-polarized but
frequency degenerate entangled modes \cite{PRL2009}, say $a$ and $b$. As
shown in Fig.~\ref{FigurePartScheme}, by means of a $\lambda /2$ wave plate
and a polarizing beam--splitter (PBS) it is possible to manipulate the
entangled state in order to obtain two \textit{independent} single-mode
squeezed vacuum states \cite{JOB2005} $a_{1}$ and $b_{1}$, with orthogonal
squeezing phases. According to the notation introduced in Section \ref%
{subSect:notation}, the correlation matrix of the latter modes can be
written as 
\begin{eqnarray}
{V}^{a_{1},b_{1}} &=& B(1/2)V_{2}^{a,b}(r^{\prime })B(1/2)  \notag \\
&=& V_{1}^{a}(r^{\prime })\oplus V_{1}^{b}(-r^{\prime }\ )\,.
\label{resource1}
\end{eqnarray}%
Then, combining the mode $b_{1}$ with the vacuum $v$ by means of a balanced
polarization insensitive beam splitter (the red plate in Fig.~\ref%
{FigurePartScheme} with $\eta =1/2$), we generate the pair of modes $%
c_{1},c_{2}$ with correlation matrix 
\begin{eqnarray}\label{eq:Vc1c2}
V^{c_{1},c_{2}} &=&B\left( \frac{1}{2}\right) [V_{1}^{b_{1}}(-r)\oplus
V_{0}]B\left( \frac{1}{2}\right) \\
&=&[S(-\frac{r}{4})\oplus S((-\frac{r}{4})]V_{2}(-\frac{r}{2})[S((-\frac{r}{4%
})\oplus S((-\frac{r}{4})]\,. \notag
\end{eqnarray}
Notice that, up to local (single-mode) operations, $c_{1}$ and $c_{2}$ are
in a TMSV state with squeezing parameter $r^{\prime }=-r/2$, \textit{i.e.} half of the original two--mode squeezing characterizing the pair $a$ and $b$ at the OPO output. At the same time, we will have at disposal an auxiliary single--mode squeezed vacuum $a_{1}$ that, in a certain sense, carries the second half of the original squeezing.

Summarizing, at the output of the above described generation stage we have at disposal three optical modes: the entangled pair $c_{1},c_{2}$,
which will play the role of the probe state for testing the
entanglement--breaking properties of the Gaussian maps $\Phi _{1}$ and $\Phi
_{2}$, and the squeezed mode $a_{1}$ which will be used as a resource for
mimicking a single-mode squeezer.
We expect that suitably setting the OPO squeezing $r$ and the beam splitters transmissivity $\eta$, we can find that the final state, of the pair $(c_1,\,c_2)$, is separable under the action of $\Phi_{1}$ and entangled for $\Phi_{2}$.

\subsection{Channel stage\label{subSect:expII}}

In this Section, we will show how to implement the maps $\Phi _{1}$ and $%
\Phi _{2}$ defined in Eq.s (\ref{phi1},\ref{phi2}), and graphically
represented in Fig.\ \ref{fig:simplifiedsetups}.

\begin{figure}[h]
\includegraphics[width=0.85\columnwidth]{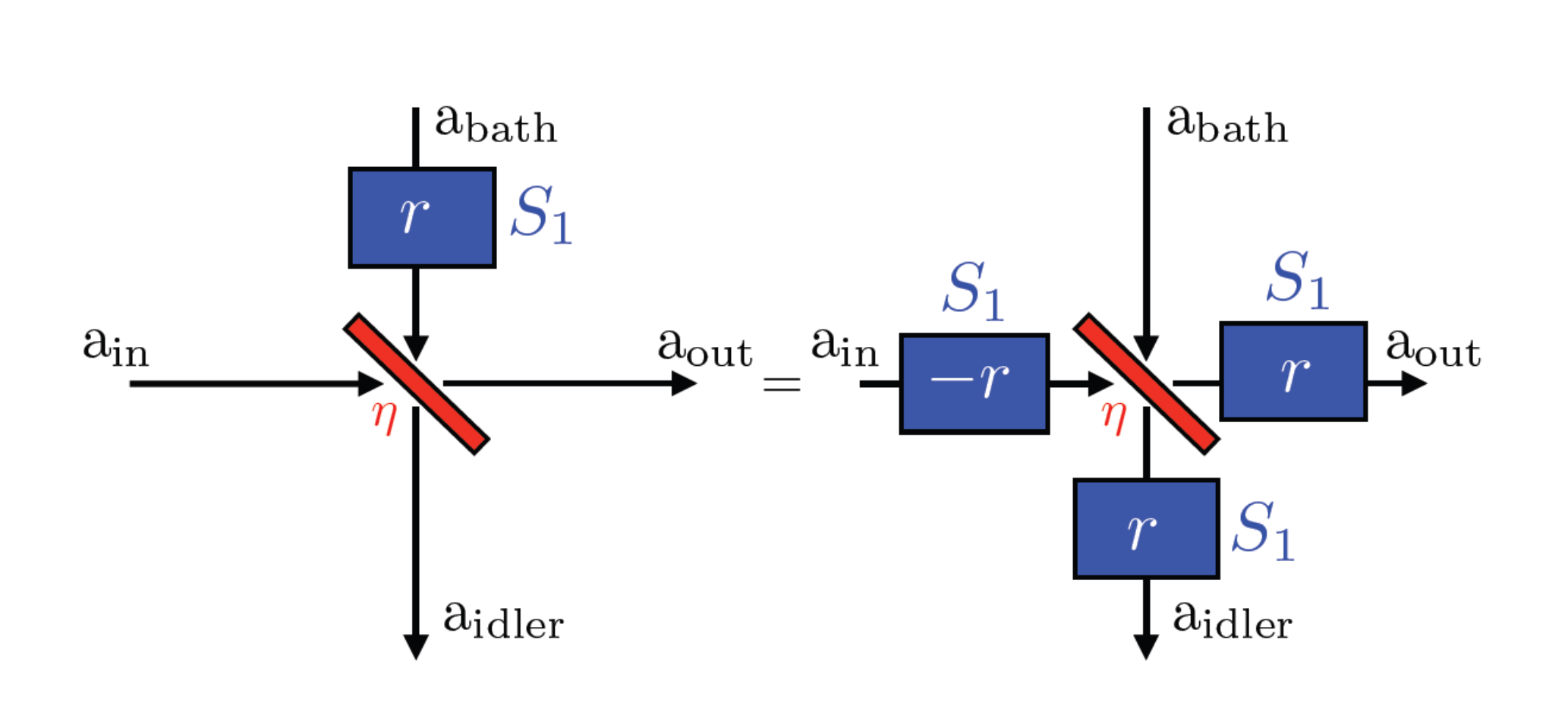}
\caption{(Color online) Graphical representation of the equivalence between
mixing a single mode squeezing and a generic state onto a beam splitter of a given
transmissivity $\eta$ and a more complex operation consisting in the sequence
$\mathcal{S}(r)\circ \Phi _{At}\circ S(-r)$. In both case the \textit{idler} mode is
traced out. This shows how a squeezed ancilla mode can be used to
effectively realize a squeezing operation on a given input state.}
\label{fig:property}
\end{figure}

Let us first consider $\Phi _{1}$. As recalled in Section \ref%
{subSect:notation}, each attenuation map $\Phi _{At}(\eta )$ can be directly
implemented by letting the incoming mode pass through a beam splitter of
transmissivity $\eta $. Less trivial is the passive implementation of the local
squeezing $\mathcal{S}(r)$ without using another OPO. This can be
indirectly achieved mixing the auxiliary squeezed mode $a_{1}$ with $c_{1}$ onto a beam splitter of transmissivity $\eta$. As a matter of fact, by observing that 
\begin{equation}
\lbrack S(r)\oplus S(r)]B(\eta )=B(\eta )[S(r)\oplus S(r)]\,,
\label{commutation}
\end{equation}
it derives that combining an incoming mode with a single mode squeezed
vacuum on a beam splitter is equivalent to indirectly attenuating and then
squeezing the system, as graphically shown in Fig.\ \ref{fig:property}.
More
precisely, we have that the effect of the first optical circuit of Fig.\ \ref%
{fig:property}, tracing out the idler mode, is equivalent to the sequence 
\begin{equation}
\mathcal{S}(r)\circ \Phi _{At}\circ S(-r).
\end{equation}%
\begin{figure}[!ht]
\begin{center}
\includegraphics[width=1\columnwidth]{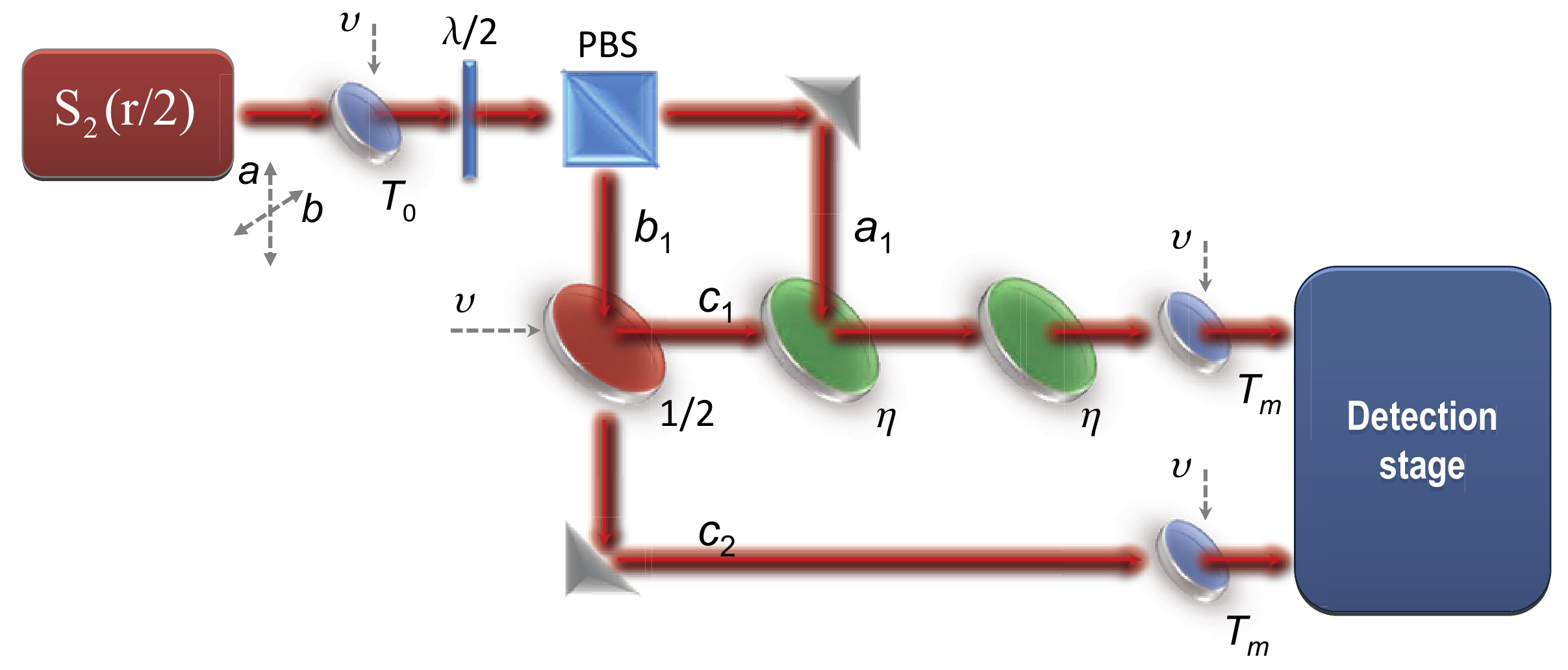}
\end{center}
\caption{(Color online) Full scheme composed by the preparation
stage (see Fig. (\protect\ref{FigurePartScheme})) followed by the
implementation of $\Phi _{1}$ and $\Phi _{2}$ by means of two BSs with
transmissivity $\protect\eta $ (lower right corner of the picture). Three
fictitious BSs, see Sec. IV A, mimic the effects of losses (BS with
transmissivity $T_{0}$) and detection inefficiencies (BSs with
transmissivity $T_{m}$).}
\label{FigureComplScheme}
\end{figure}
By applying this equivalence property, we can easily implement the action of 
$\Phi _{1}$ on the incoming mode $c_{1}$, as pictorially represented in the
full experimental set--up given in Fig. \ref{FigureComplScheme}. Here the green
beam splitters act on the incoming mode $c_{1}$ as 
\begin{equation}
\Phi _{At}\circ \mathcal{S}(r)\circ \Phi _{At}\circ S(-r)=\Phi _{1}\circ
S(-r)\, ,  \label{green}
\end{equation}
thus experimentally realizing the map of Eq.(\ref{phi1}) up to the unitary
transformation $S(-r)$. We recall that the entanglement-breaking properties
of a map, are invariant under unitary redefinition of the input and output
spaces~\cite{EBT}, that is $\Phi _{1}\circ S(-r)\in \mathrm{EB}$ iff $\Phi _{1}\in 
\mathrm{EB}$.

On the other hand, the implementation of the channel $\Phi_2$ defined in
Eq.\ (\ref{phi2}) can be straightforwardly implemented by discarding the
auxiliary mode $a_1$ and substituting it with the vacuum. In other words,
one should simply let the mode $c_1$ pass through the green beam splitters
without feeding any light in the empty ports.

We can therefore conclude that the experimental setup represented in Fig.\ ~\ref{FigureComplScheme} is, up to experimental losses, equivalent to the
theoretical scheme in Fig.\ \ref{fig:simplifiedsetups}. For the sake of
clearness, let us point out that while in the theoretical scheme of Fig.~\ref{fig:simplifiedsetups} 
the squeezing parameters $r^{\prime }$ and $r$ are totally independent, in the 
realistic setup of Fig.~\ref{FigureComplScheme} the structure of the scheme forces 
$r^{\prime }=-r/2$. Nonetheless, this lack of freedom does not affect the
feasibility of the experiment.

\section{Effective channel properties\label{Sect:Results}}

As explained in Section~\ref{subSect:teoscheme} (see Eq.s (\ref{amend3}-\ref%
{phi2})), in order to experimentally prove the existence of Gaussian
amendable channels we need to show that $\Phi _{1}$ is entanglement-breaking
while $\Phi _{2}$ is not. This can be done by measuring the output state of
our experimental circuit and checking its separability for different choices
of the experimental parameters. In particular we will apply the PPT criterion \cite{PPT01,PPT02,PPT03} to the CM of the output state.

Here we discuss the proposed experimental scheme and we theoretically
estimate $V^{out}$, the expected CM for the final state. By
writing it in $2\times 2$ blocks 
\begin{equation}
V^{out}=%
\begin{pmatrix}
\mathbf{A} & \mathbf{C} \\ 
\mathbf{C^{\top }} & \mathbf{B}%
\end{pmatrix}%
\, ,
\end{equation}%
one can easily compute the minimum symplectic eigenvalue $\nu $ of the partially transposed state 
\begin{equation}
\nu =\sqrt{\frac{\Sigma -\sqrt{\Sigma ^{2}-4\det [V^{out}]}}{2}}
\label{eq:nu}
\end{equation}%
where $\Sigma =\det [\mathbf{A}]+\det [\mathbf{B}]-2\det [\mathbf{C}]$. From
the PPT criterion it can be shown that the output state is
entangled if and only if 
\begin{equation}
\nu ^{2}<\frac{1}{4}  \label{criterion}
\end{equation}%
(see \cite{gaussian03} and references therein). The relation above, provides
a necessary and sufficient criterion for testing the separability of the
output state and thus for studying the entanglement-breaking properties of
the applied map.

\begin{figure}[th]
\centering {\includegraphics[width=0.7\columnwidth]{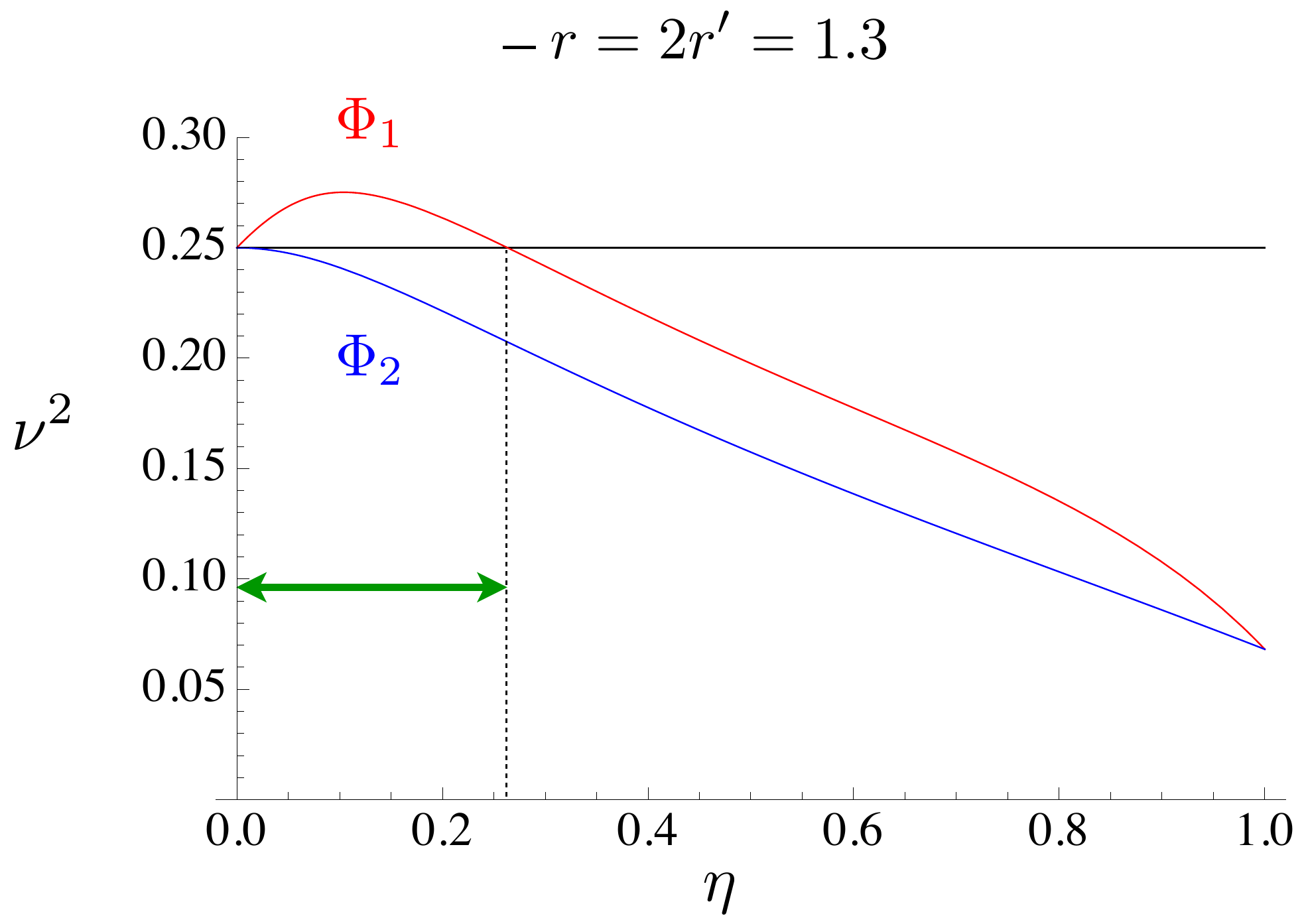}}
\caption{(Color online) Entanglement witness parameter $\protect\nu ^{2}$ as
a function of the transmissivity $\protect\eta $, computed for the outcomes
of $\Phi _{1}$ and $\Phi _{2}$, in absence of losses and noise. As expected, 
$\Phi _{2}$ preserves the entanglement of the incoming twin-beam for all $%
\protect\eta $, indeed $\protect\nu ^{2}<1/4$ (as signaled by
the lower (blue) curve). On the other hand, there exists a finite interval of
transmissivity such that the output state of $\Phi _{1}$ is separable
($\protect\nu ^{2}>1/4$).}
\label{fig:nuquadro2}
\end{figure}

The behaviour of $\nu^{2}(\eta)$ for initial squeezing $\left| r \right|=1.3$ is given in Fig.~\ref{fig:nuquadro2}, that refers to the case of ideal (lossless) preparation stage and
detectors with unit efficiency. It results that, if on the one hand $\Phi
_{2}$ can never become ${\mathrm{EB}}$ for any value of the transmissivity $%
\eta$ (\textit{i.e.} $\nu ^{2}$ is always lower than $1/4$ so that the state keeps its entanglement), on the other
hand there exists a finite interval of $\eta$ such that $\Phi _{1}\in {%
\mathrm{EB}}$ and thus its output state is separable. This separability
interval has been computed in~\cite{amendableChTEO} and corresponds to $\eta
\leq \tilde{\eta}(r^{\prime })$ with 
\begin{equation}
\tilde{\eta}(r^{\prime })=\frac{1}{2}\left( \cosh (2r^{\prime })-\sqrt{%
2\cosh (2r^{\prime })-1}\right) \text{csch}^{2}(r^{\prime })\,.
\end{equation}%
In the next subsection, we will consider the effects of losses, detection
inefficiencies and measurement uncertainties. 

\subsection{Measurement uncertainty, losses and inefficiencies\label{subSect:losses}}

In a realistic implementation we cannot neglect the statistical uncertainty
affecting the measurement process and, at the same time, we also have to
consider the effects of losses (decoherence) and detection efficiency.

In order to take into account the experimental indeterminacy into Eq.
(\ref{eq:nu}) we have considered typical experimental values for the
uncertainties relative to the CM elements. These values are used in
propagating the measurement's errors into the formula that gives $\nu ^{2}$
in terms of CM elements (a detailed discussion on the errors affecting the
different elements can be found in Ref. \cite{JOSAB2010}). Thus we obtain
the statistical error $\delta (\nu ^{2})$ for $\nu ^{2}$. From the
experimental point of view claiming that $\Phi _{1}\in {\mathrm{EB}} $
requires that $\nu _{\Phi _{1}}^{2}-1/4>2\delta (\nu ^{2})$, \textit{i.e.}\
the distance from the separability threshold must overcome the measurement
confidence interval (\textit{i.e.} twice the uncertainty).

\begin{figure}[th]
\centering
\subfigure[]{\label{fig:err05}
\includegraphics[width=0.6\columnwidth]{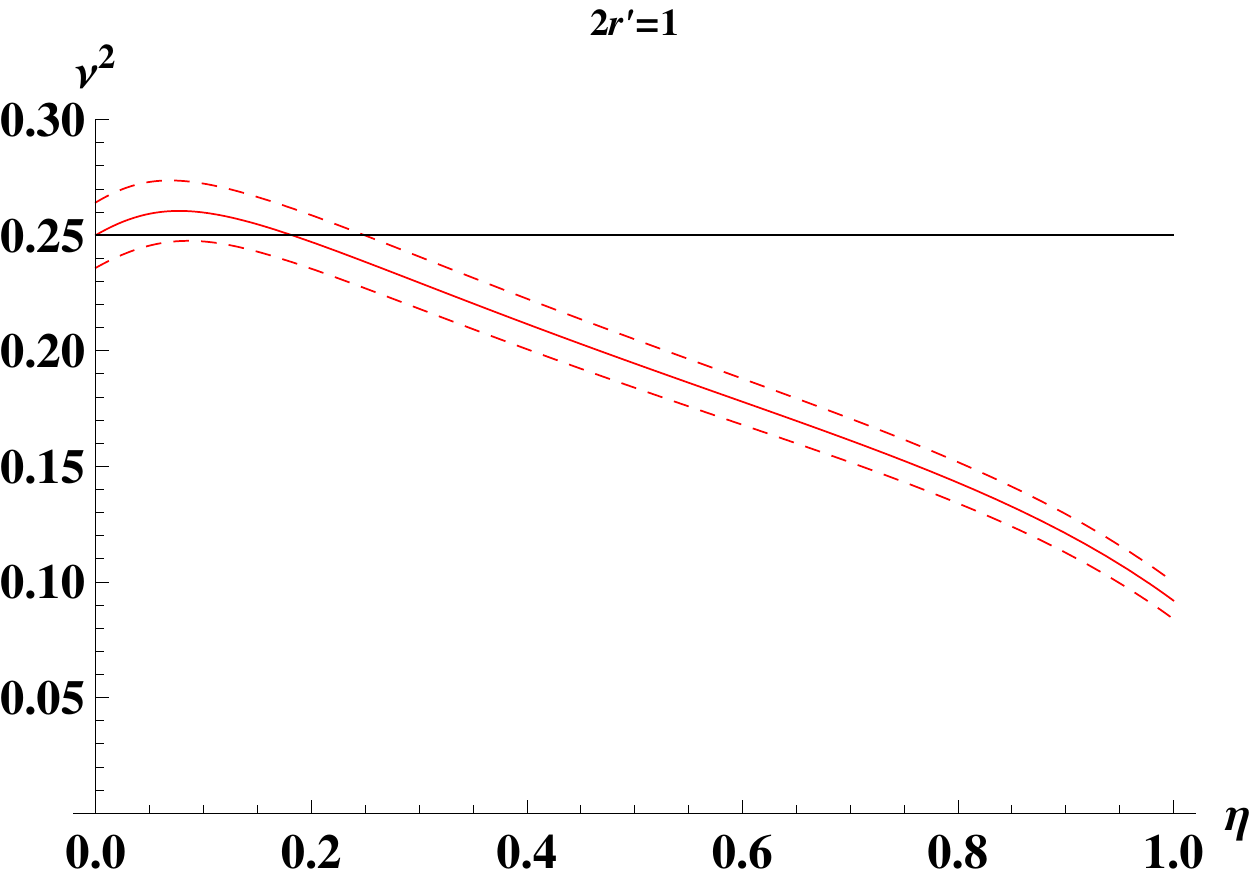}} \hspace{5mm} %
\subfigure[] {\label{fig:err065}\includegraphics[width=0.6%
\columnwidth]{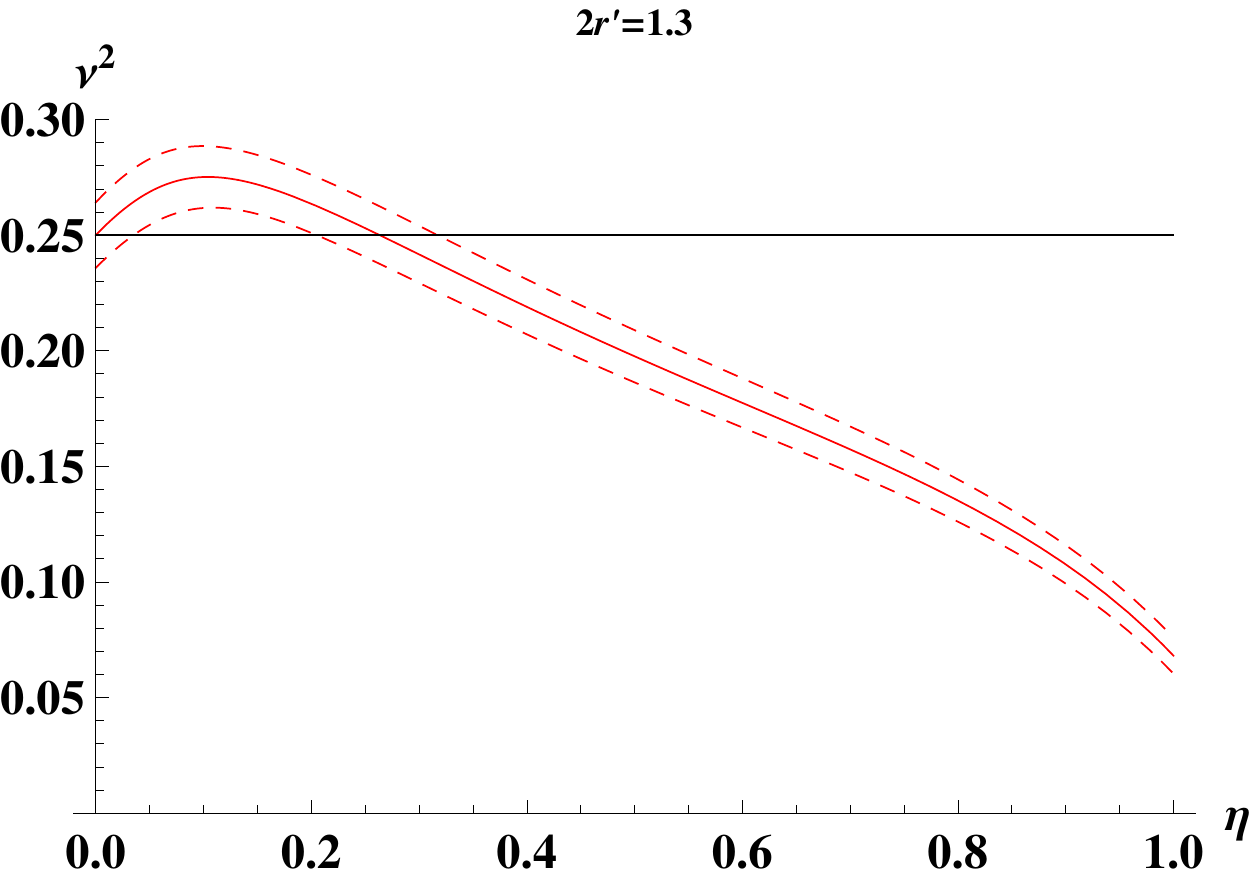}}
\caption{(Color online)Entanglement witness parameter $\protect\nu ^{2}$ as
a function of $\protect\eta $ for two different values of the squeezing
parameter associated to the initial modes $a$ and $b$: $r^{\prime }=-r/2=0.5$
in (a) and $r^{\prime }=-r/2=0.65$ in (b). The dashed lines indicate the
confidence interval one should expect for $\protect\nu ^{2}$ in a typical
measurement of the covariance matrix elements via homodyne detection.}
\label{fig:6}
\end{figure}

In Fig.~\ref{fig:err05} we have fixed $r^{\prime}=-r/2=0.5$ and plotted $\nu
^{2}$ as a function of $\eta$. The dashed lines represent the boundaries of
the confidence interval for $\nu ^{2}$. 
From this plot we conclude that in this case the confidence interval $2\delta (\nu ^{2})$
would make ambiguous, from the experimental point of view, the statement
that $\Phi _{1}\in {\mathrm{EB}}$. This ambiguity can be overcome by
considering an increased level of the squeezing for the pure state generated
by the type--II\ OPO. For example, it is sufficient to raise $|r|$ from $1$
to $1.3$ to obtained a clear experimental proof that $\Phi _{1}\in {%
\mathrm{EB}}$ for $\eta \leq \tilde{\eta}$, as shown in Fig. \ref{fig:err065}.
Here $\nu ^{2}\left( \eta \right) $ is plotted for $r^{\prime}=-r/2=0.65$,
and $\nu _{\Phi _{1}}^{2}-1/4$ is greater than the expected confidence
interval in a range of values for $\eta$ contained in $[0,\tilde{\eta}]$.

Furthermore, real experiments face the effects of absorption losses and
non--ideal detection, which can be modeled by the three fictitious beam
splitters we have introduced in Fig.~\ref{FigureComplScheme}. The first one
of transmissivity $T_{0}$ (on the left) simulates the effects of losses and in particular, the OPO cavity escape efficiency \cite{PRL2009} that unavoidably makes any state at the output of an OPO cavity a mixed one \cite{mixed}.
The last two beam splitters of transmissivity $T_{m}$ (on the right)
model the inefficiency of the detectors.

\begin{figure}[th]
\centering {\includegraphics[width=0.6\columnwidth]{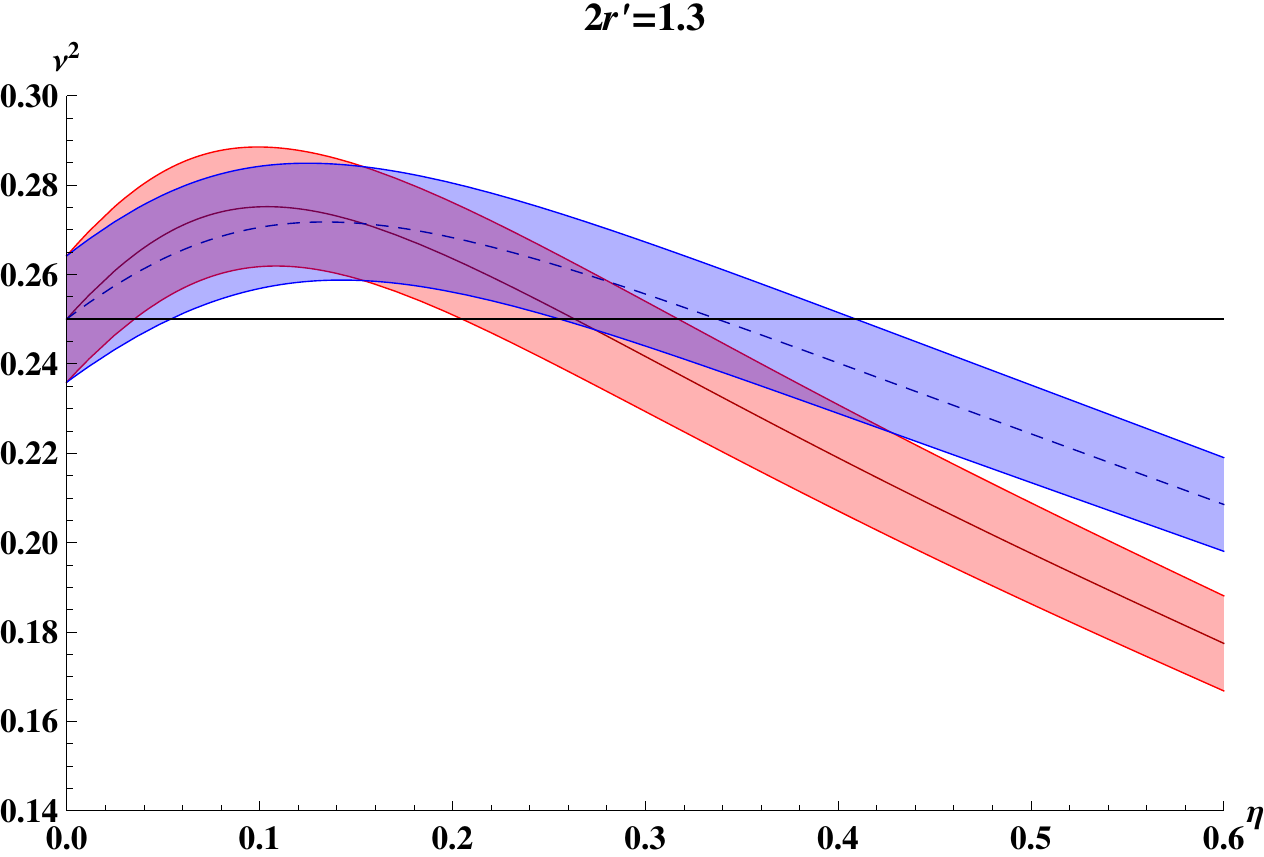}}\hspace{%
15mm}
\caption{(Color online) Entanglement witness parameter $\protect\nu ^{2}$ as
a function of $\protect\eta $. We have fixed the squeezing parameter
associated to the initial modes $a$ and $b$ to $r^{\prime }=-r/2=0.65$.
We compare the ideal case (no-loss and unit detection efficiency, red line)
with a realistic case where
$T_{0}=0.75$ and $T_{m}=0.90$ (blu dashed line). The plotted lines
correspond to the expectation values while shadowed areas encompass confidence
intervals. The plot, clearly, shows that the proposed scheme is quite
insensitive to losses and detection inefficiency.}
\label{fig:real065}
\end{figure}

It is interesting to see that the conclusions retrieved from the analysis
performed in Fig.~\ref{fig:err065} are still valid if losses and detection
inefficiencies are taken into account.
In Fig.~\ref{fig:real065} we plot the behavior of $\nu
^{2}\left( \eta \right) $ in a realistic scenario, setting the losses at $25\%$ so that $%
T_{0}=0.75$ and detection efficiency at $T_{m}=0.90$, and assuming the same
statistical indeterminacy used in the case without losses.
The effect of $T_{0}<1$ and $T_{m}<1$
is, on one hand, to reduce the maximum value for $\nu ^{2}$ inside the EB
region (the maximum also moves to a higher $\eta$'s value), 
on the other hand, to enlarge the
$\eta $ interval for which $\Phi _{1}=\Phi ^{2}\in {\mathrm{EB}}$. 

We note that
while reducing the weight of losses and detection inefficiencies is surely
possible ($T_{0}=0.95$ and $T_{m}=0.97$ have been recently reported \cite%
{Schnabel2013}) experimental indeterminacy cannot be avoided and, as far as
we know, the value used in Ref. \cite{JOSAB2010} is the lowest one for the
experimental determination of the CM of a bipartite Gaussian state.

\section{Conclusions}

In this work we have proposed a realistic quantum optics experiment based
on continuous variable systems that would provide the existence of Gaussian 
amendable maps and give more insight on entanglement breaking
channels from a practical point of view. 

The proposed scheme is translated into a rather simple experimental set--up. Indeed, it is based on a single initial
non-linear operation (realized by a type--II sub--threshold OPO) which has the role of preparing both the input entangled state and a squeezed ancilla.
The rest of the scheme is extremely simple since it requires only passive
operations such as beam splitters and wave-plates.

The proposal has been realistically analyzed by taking into account the
typical statistical uncertainty of Gaussian state quantum homodyne tomography. The
effects of losses and detector inefficiencies have also been considered. We have
shown that, even in presence of such errors and losses, the experiment
is still feasible. Indeed, a conclusive test can be achieved by
appropriately tuning the experimental parameters.
The proposed scheme can be readily implemented in any laboratory having at 
disposal a running source of bipartite Gaussian entangled states.

\end{document}